# Air cold atmospheric plasma with patterns for anaplastic squamous cell carcinoma treatment

Fan Bai[1,2], Yingjie Lu[2], Yujie Zhi[3], Yueye Huang[2], Long Li[1,2], Jiaoxiao Luo[4], Jamoliddin Razzokov[5], Olga Koval[6], Maksudbek Yusupov[7], Guojun Chen[8,9], Zhitong Chen[1,2]*

[1] Institute of Biomedical and Health Engineering, Shenzhen Institute of Advanced Technology, Chinese Academy of Sciences, Shenzhen 518055, China

[2] National Innovation Center for Advanced Medical Devices, Shenzhen, 518000, China

[3] China electronics standardization institute, Beijing 100007, China

[4] Institute of Mechanics, Chinese Academy of Sciences, Beijing 100190, China

[5] Institute of Material Sciences, Academy of Sciences of the Republic of Uzbekistan, Chingiz Aytmatov 2b, Tashkent 100084, Uzbekistan

[6] Institute of Chemical Biology and Fundamental Medicine, Siberian Branch of the Russian Academy of Sciences, Akad. Lavrentiev Ave. 8, Novosibirsk 630090, Russia

[7] Arifov Institute of Ion-Plasma and Laser Technologies, Academy of Sciences of the Republic of Uzbekistan, Durmon yuli 33, Tashkent 100125, Uzbekistan

[8] Department of Biomedical Engineering, McGill University, Montreal, Quebec H3A 1A3, Canada

[9] Rosalind & Morris Goodman Cancer Institute, McGill University, Montreal, Quebec H3A 1A3, Canada

* To whom correspondence may be addressed: zt.chen@nmed.org.cn (Z.C.)

## Abstract

In recent years, cold atmospheric plasma (CAP) using inert gas has been successfully applied for biomedicine, such as sterilization, wound healing, skin diseases, and tumor treatment. Here, we reported air cold atmospheric plasma with three different patterns (I. Non: basic square grid structure; II. Square: basic square grid structure + square node; III. Circle: basic square grid structure + circle node) for anaplastic squamous cell carcinoma treatment (VX2 cell line). Various plasma diagnostic techniques were applied to evaluate the physics of air CAP with patterns such as discharge voltage, plasma initial generating process, plasma temperature, and optical emission spectroscopy (OES). The direct effects of air CAP with patterns on anaplastic squamous cell carcinoma treatment (VX2 cell line) were investigated in vitro. We also studied the ROS (reactive oxygen species) and RNS (reactive nitrogen species) generation in cultured media released from VX2 cells after the treatment of air CAP with patterns. The results showed that the air CAP with circle-pattern generated more active substances during at 60s treatment time, which resulted in a higher death rate of VX2 cells. These initial observations establish the air CAP with patterns as potential clinical applications for cancer therapy.

# 1. Introduction

Cold atmospheric plasma (CAP) is a near-room temperature plasma mainly composed of reactive oxygen species (ROS), reactive nitrogen species (RNS), free radicals, UV photons, charged particles, and electric fields[1]. CAP has been widely used in the biomedical fields over the past decade, such as sterilization, wound healing, and skin disease treatment, especially in the treatment of cancer[2-8]. It has been reported to be effective in killing various tumor cells, including lung cancer, liver cancer, skin cancer, breast cancer, cervical cancer, and brain cancer[9-13]. For example, Liu's lab developed a fillable plasma-activated biogel for local post-operative treatment of cancer[14]. Their results indicate the plasma-activated biogel eliminates residual tumors after surgery and inhibits in situ recurrence without evident systemic toxicity. Bekeschus et al. showed CAP treatment significantly reduced the metabolic activity of Mia Paca-2 and PAN-1 in vitro cultures, correlating with reduced cell viability and that PAN-1 exhibits higher resilience to CAP treatment (6% metabolic activity)[15]. CAP is known to have limited tissue penetration[16, 17]. To overcome this challenge, Chen et al. developed a hollow-structured microneedle patch (hMN) to deliver CAP through the skin into the tumor tissue[18]. Their results indicated the CAP/hMN device significantly induced an enhanced anti-tumor effect. The delivery platform has also been combined with immunotherapeutics (i.e., immune checkpoint inhibitors) to further enhance the antitumor efficacy. The efficacy of CAP in proposed applications relies on the synergistic action of reactive nitrogen species (RNS), reactive oxygen species (ROS), UV photons, free radicals, charged particles, and electric fields[19-22].

So far, most cancer treatments via plasma utilize inert gases, which limits plasma for cancer clinical applications[23-26]. Here, we proposed air CAP with three patterns applied to anaplastic squamous cell carcinoma treatment (VX2 cell line). The effect of air CAP with three patterns on anaplastic squamous cell carcinoma in vitro was evaluated, which would promote the range of plasma clinical applications, especially for cancer treatment.

## 2. Materials and Methods

### 2.1 Experimental Device Configuration

As shown in Fig. 1, the air CAP with patterns device consists of a high voltage electrode, a copper mesh ground electrode, and a polytetrafluoroethylene (PEFT) plate sandwiched between the two electrodes, all with a thickness of 1mm. Uniform plasma can be generated by AC high voltage power supply. Three different patterns were designed: I. Non: basic square grid structure; II. Square: basic square grid structure + square node; III. Circle: basic square grid structure + circle node. The total length of the copper mesh ground electrode of Non/Square/Circle in the metal ring is 142mm/172.6mm/129mm. The surface temperature of air CAP with patterns slightly increases from room temperature to about 36 °C and tends to be stable after 60s of action. The constant frequency was 10 kHz and the peak-peak voltage was 8kV. The corresponding peak-peak discharge voltage and frequency were 4.2kV/3.8kV/4.18kV and 16.8 kHz/17kHz/17.8kHz separately for air CAP with non-pattern/ square-pattern/ circle-pattern.

### 2.2 Optical Emission Spectroscopy (OES) Spectra Measurement

We used a UV/visible spectrometer (Maya Pro 2000, Ocean Optics, China) to measure the OES of air CAP with three patterns in the wavelength range of 200 ~ 800 nm. The optical probe was directly placed 1cm away from the discharge area to detect the surface CAP discharge spectra in the cell mesh enclosed by copper wires.

### 2.3 Cell Culture

Anaplastic squamous cell carcinoma (VX2 cell line, Meisen CTCC, Zhejiang, China) was cultured in Dulbecco's Modified Eagle's Medium (DMEM, Solarbio, Beijing, China) supplemented with 10% fetal bovine serum (Sorlarbio) and 1% penicillin-streptomycin (Solarbio). Cultures were maintained at 37°C in a cell incubator containing 5%$CO_2$. VX2 cell lines were plated in 6-well plates at a density of $1\times10^6$ cells per well in 2mL of complete culture medium for subsequent plasma experiments.

### 2.4 The treatment procedure of air CAP with patterns

The general experimental procedure is shown in Fig. 2. After Removing the medium supernatant, VX2 cell lines were treated with air CAP with three patterns of 0s, 10s, 20s, 30s, 60s, and 120s in turn. The same treatment times were applied to DMEM treated by air CAP with patterns.

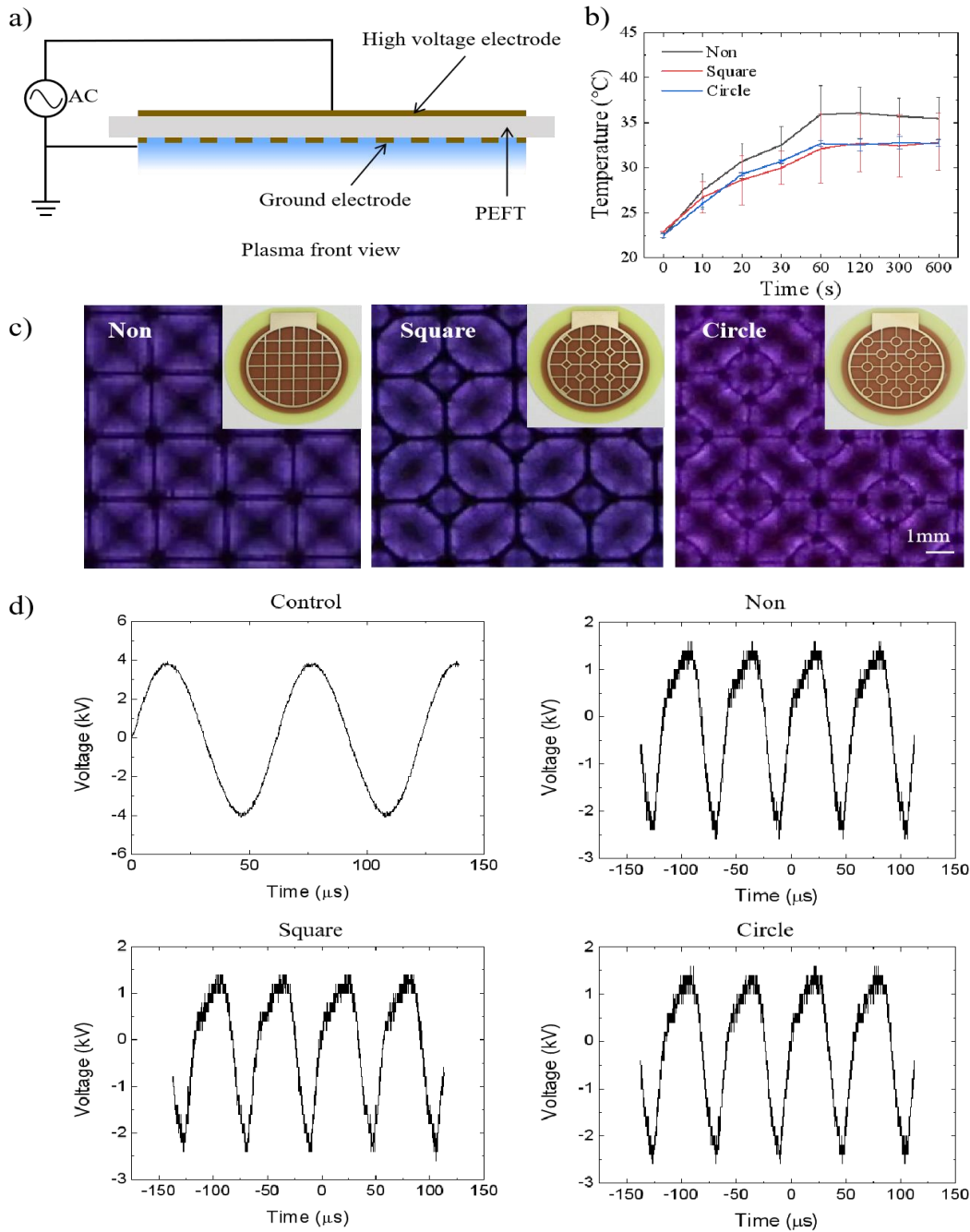

Fig. 1. a) Schematic representation of air CAP with patterns (I. Non: basic square grid structure; II. Square: basic square grid structure + square node; III. Circle: basic square grid structure + circle node); b) Temperature evolution of air CAP with three different patterns over time; c) Air CAP with Non/Square/Circle-patterns generate different uniform surface plasma; d) Time evolution of the typical voltage waveform of air CAP with three patterns during the plasma treatment.

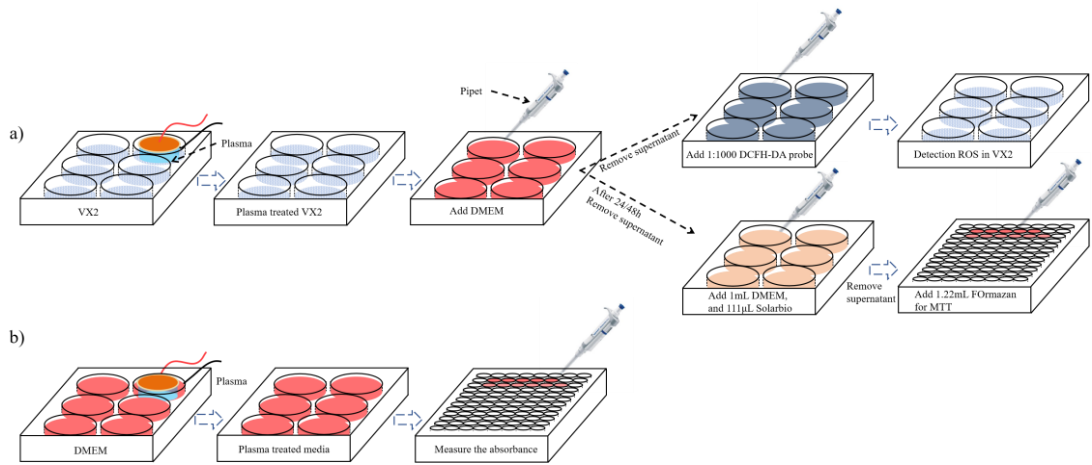

Fig. 2. The flow diagram of air CAP with patterns applied to a) VX2 cell line or b) DMEM. a) Experimental procedure of ROS and cell viability determination (up) and NO determination (down) in VX2 cell line. b) Experimental procedure of NO and $H_2O_2$ concentration determination in DMEM.

## 2.5 Determination of ROS Concentration

Based on the manufacturer's protocol, the ROS detection kit (Beyotime, Shanghai, China) was used to measure the concentration of ROS in cells. After CAP treatment, 1mL complete culture medium was added to each well to incubate for 0min, 3min, 30min, 60min, and 120min. Then the supernatant was removed and a 1mL DCFH-DA probe (serum-free medium at a 1:1000 ratio) was added to each well. The cells were incubated in a cell incubator for 20min. After the incubation, the cells were washed three times with PBS buffer (Solarbio) to adequately remove the DCFH-DA probe that did not enter the cells. The fluorescence intensity of each well was measured with a microplate reader at Ex/Em: 488/525 nm.

## 2.6 Determination of NO Concentration

NO detection kit (Beyotime) was used to detect RNS levels in cells. First, lysis cells with cell lysate (Beyotime) for 10-15 minutes, centrifuge for 3-5 minutes to take the supernatant for subsequent nitric oxide detection. 50 μL of samples and 50 μL of standard $NaNO_2$ solution were added to a 96-well plate. Then, 50 μL of Griess Reagent I and 50 μL of Griess Reagent II were added to each well. The absorbance was measured at 540 nm with a microplate reader.

## 2.7 Cell Viability Following Air CAP with Patterns Treatment

The methyl thiazolyl diphenyl-tetrazolium bromide (MTT) assay (Solarbio) was used to evaluate cell viability. After CAP treatment, cells were incubated in a cell incubator at 37 °C for 24h and 48h. At the end of each incubation time point, the supernatant was discarded, and 1mL of fresh medium and 111 μL of MTT solution (Solarbio) were added to each well for a 4h-incubation time. The supernatant was then discarded and 1.22 mL of Formazan solution (Solarbio) was added to each well to dissolve the dark purple product Formazan. After the crystals were fully dissolved, the absorbance of each well was measured at 490 nm in a microplate reader.

### 2.8 Determination of $H_2O_2$ Concentration in Medium

A hydrogen peroxide assay kit (Beyotime) was used to detect the concentration of $H_2O_2$ in plasma-treated mediums based on the instructions provided by the manufacturer. Briefly, 50 μL of samples and 50 μL of standard $H_2O_2$ solution were added to a 96-well plate, and then 100 μL of hydrogen peroxide test reagent was added to each well. After 30 minutes of gentle mixing, The absorbance was measured at 560 nm with a microplate reader.

### 2.9 Determination of NO Concentration in Medium

NO detection kit (Beyotime) was used to measure the amount of NO in plasma-treated mediums. According to the protocol of the manufacturer, 50 μL of samples and 50 μL of standard $NaNO_2$ solution were added to a 96-well plate. Then, 50 μL of Griess Reagent I and 50 μL of Griess Reagent II were added to each well. The absorbance was measured at 540 nm with a microplate reader.

### 2.10 Definition of Control and Statistical Analysis

In Figures 6, 7, 8, and 10, "0 s" treatments represented no CAP treatment. In Fig. 9, "control" groups represented no CAP treatment. Thus, "0s" and "control" were used as control. Results were plotted using Origin 2022 as mean ± standard deviation.

# 3. Results

## 3.1. The optical emission spectra (OES) of air CAP with patterns

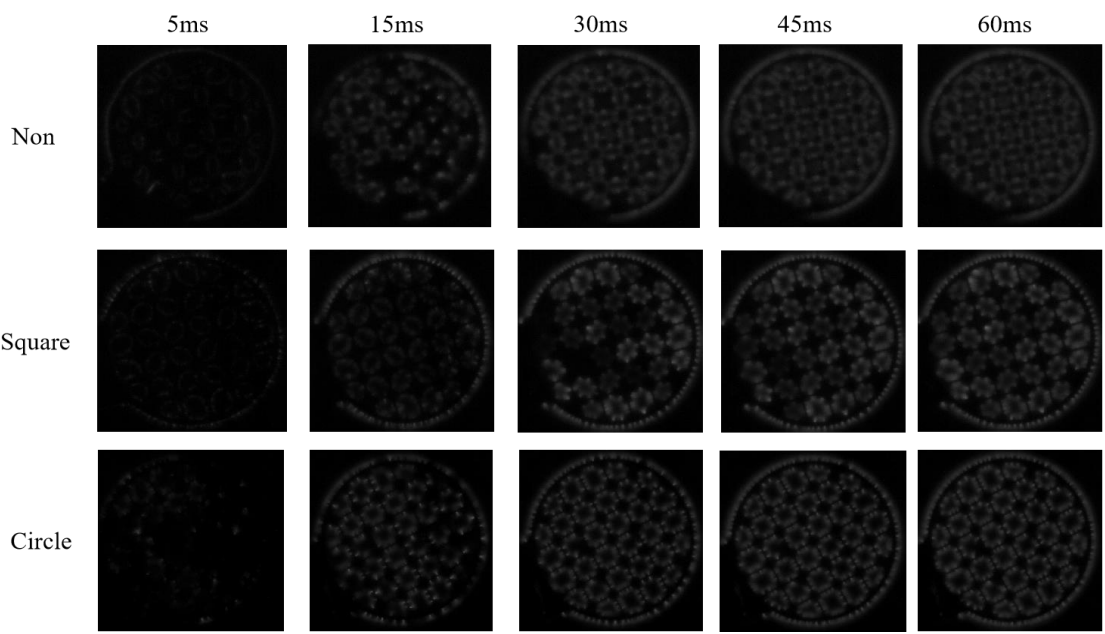

Fig. 3. The excitation process of plasma with three patterns: Non, Suqare, and Circle. The missing part on the left in the picture is connected to the wire.

Fig.3 shows the image of the air CAP device at different times after the power was turned on. From the graph, it can be observed that all three patterns took about 60ms to reach a steady state from initial excitation. The outer ring started to generate plasma at 5 ms after the power was turned on. The images taken at 15 ms intervals showed that the inner part of the ring gradually began to generate plasma. In the case of non- and circle-pattern, parts were luminated at 15ms. The whole of the non- and circle- patterns began to luminate after 15ms and finally reached a steady state at 45ms and 30 ms, respectively. While in the square-pattern, at 15ms after the power was turned on, only a small region of the pattern was luminated, and it kept flushing in the next 30ms. The square-pattern with square reached a steady state at 60ms. Overall, The circle-pattern was the first to create a uniform and stable plasma at 30 ms, the non-pattern generated a uniform plasma at 45 ms, and the square-pattern was the slowest at 60 ms.

The OES of air CAP with three patterns were shown in Fig. 4. In 1976, Pearse et al. proposed the identification method of emission line and band[27]. Most peaks appear in the UV and violet regions from 315 nm to 430 nm. The peaks in the figure indicated the presence of the key chemical element reactive nitrogen (RNS) in plasma. It can be found that none-pattern had a higher peak, but circle-pattern and square-pattern identified similar peaks. As a representative peak, the peak at 281 nm represents a decrease in photons emitted by excitation NO from $A^2\Sigma^+$ to $X^2\Pi$. The wavelength of 309 nm is defined as OH. The peak value at 337 nm represents the photon emission intensity when $N_2$ is excited from the states $C^3\Pi_u$ to $B^3\Pi_g$. The band of 405.5 nm between 400 and 430 nm can be defined as $N_2$. Species with a wavelength of 777 nm can be defined as O. These emission bands could be coming from the surrounding air.

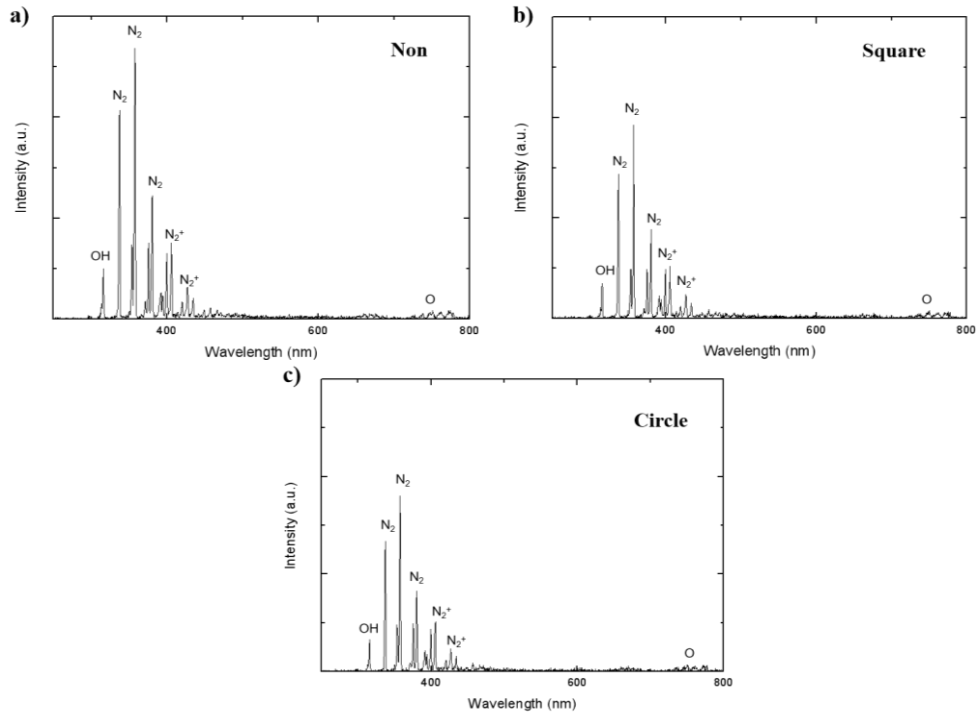

Fig. 4. Spectra of the air CAP with three patterns: a) non-pattern; b) square-pattern; c) circle-pattern.

The optical emission distributions of the air CAP with three patterns are shown in Fig. 5. The three curves of different patterns are close to each other in the middle and have the highest intensity. For the area to its left and right, the intensity decreases with the deviation of the distance. Non-pattern has the highest overall intensity, while circle-pattern has the lowest overall intensity. For the 337 nm $N_2$, and 405.5 nm $N_2^+$ emission peaks, the non-pattern and square-pattern have almost the same trend, with non-pattern having a higher intensity than square-pattern. The change rate of the circle-pattern is significantly larger, mainly due to the contribution of the electric field strength. OH and O released by circle-pattern are higher than those released by non-pattern and square-pattern, especially OES of 777 nm O. Therefore, it seems possible to manipulate the chemical species by changing the electric field through the metallic pattern on the DBD electrode sheet. For example, when air CAP with patterns is irradiated to cells, the amount of excited $N_2$ and $N_2^+$ can be increased if we use a dense metal pattern.

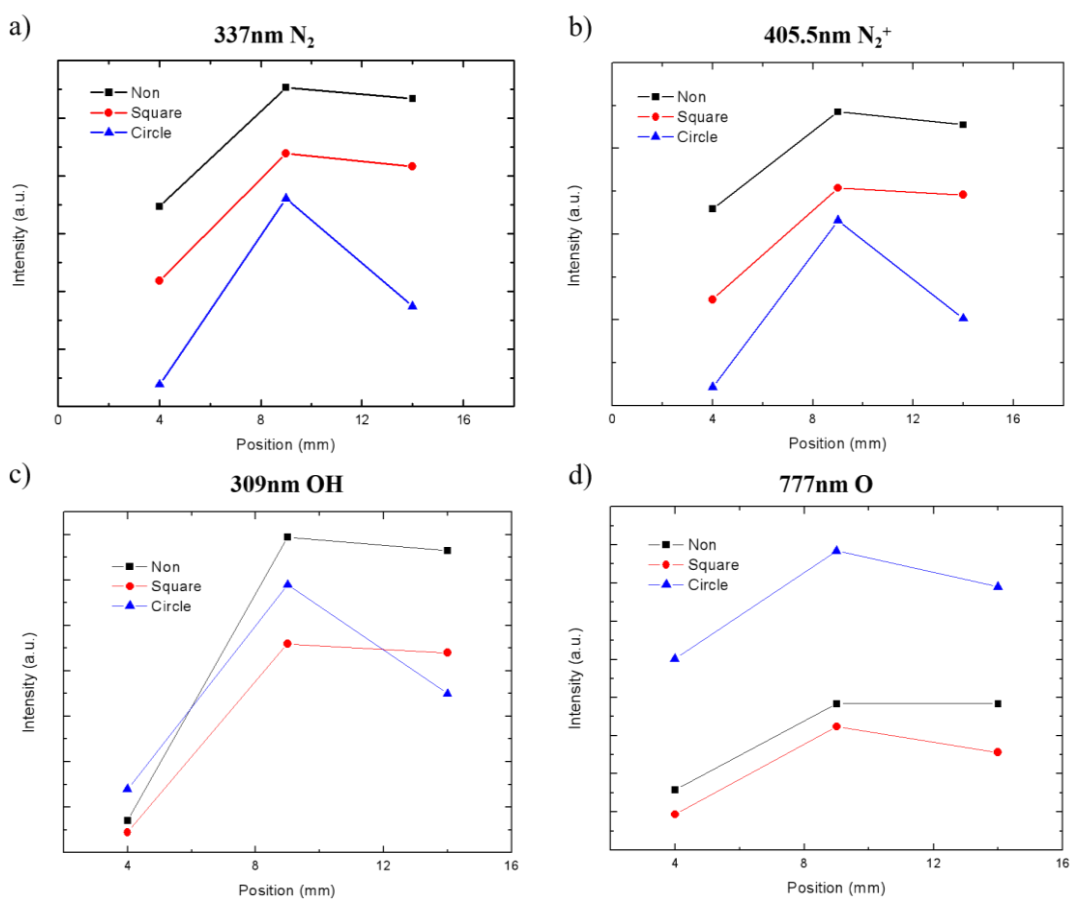

Fig. 5. Spectra of the air CAP with patterns for 337 nm $N_2$, 405.5 nm $N_2^+$, 309 nm OH, and 777nm O.

**3.2 Detection of NO and $H_2O_2$ produced by air CAP with patterns in medium**

Fig. 6a showed that $H_2O_2$ and NO concentration in the CAP-treated DMEM increased with the increase of CAP treatment time. The overall growth trends of $H_2O_2$ and NO concentration for square-pattern and circle-pattern were higher than that of non-pattern. Specifically, the growth trends for square-pattern increased rapidly after 60s and reached the maximum value at 120s, which was higher than that of non-pattern and circle-pattern. The experimental results showed that the active species induced by CAP were able to enter the solution and continue to grow with the increasing treatment time.

Fig. 6b and 6c represented the $H_2O_2$ growth rate and NO concentration in cultured media released from VX2 cells after the treatment of air CAP with patterns for 24h/48h-incubation time. It can be seen that the $H_2O_2$ growth rate and NO concentration showed a turning and decreasing trend in 60s after the continuous increase from 0s to 30s. And the $H_2O_2$ growth rate of square-pattern was higher than non-pattern and circle-pattern for 24h and 48h incubation time. In particular, the $H_2O_2$ growth rate in the cultured media after culturing for 48h dropped to 0 at 120s. In Fig. 6b, the NO concentrations of the three patterns were not significantly different within 60s, and the NO concentration of circle-pattern was significantly higher than that of the other two patterns in 60s and 120s. While the NO concentration of non-pattern was higher than that of square-pattern and circle-pattern in Fig. 6c.

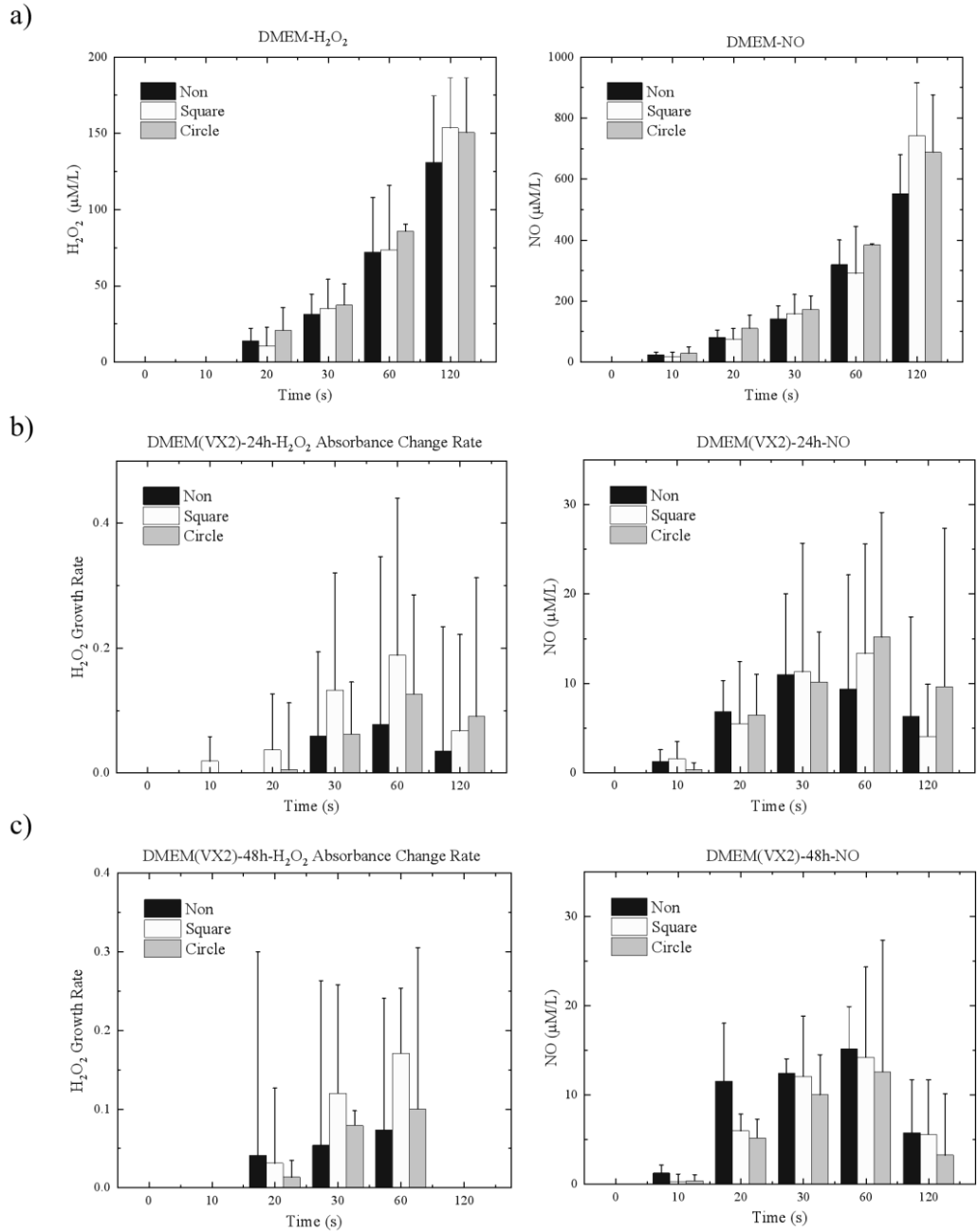

Fig. 6. a) $H_2O_2$ and NO concentration in air CAP with three patterns-treated DMEM. b) $H_2O_2$ growth rate and NO concentration in cultured media released from VX2 cells after the treatment of air CAP with patterns for 24h-incubation time. c) $H_2O_2$ growth rate and NO concentration in cultured media released from VX2 cells after the treatment of air CAP with patterns for 48h-incubation time.

### 3.3 The effect of air CAP with patterns on VX2 cell lines

### 3.3.1 Detection of NO produced by air CAP with patterns in cells

As shown in Fig. 7, the NO concentrations generated by the air CAP with three patterns increased from 0s to 30s, while the NO concentration of non-pattern and circle-pattern decreased from 60s to 120s. The square-pattern NO concentration first increased after 30s, reached the highest value in 60s, and then decreased slowly.

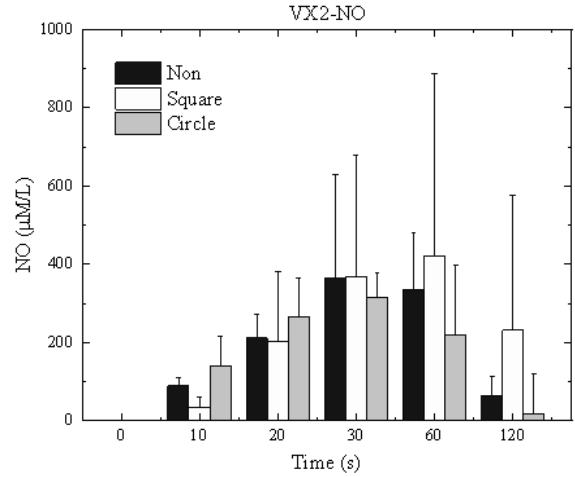

Fig. 7. NO concentration in air CAP with three patterns-treated VX2 cell lines.

### 3.3.2 Detection of ROS produced by air CAP with patterns in cells

The ROS assessment results of cytofluorimetry showed that ROS generated by both non-pattern and square-pattern reached the maximum value when CAP was treated for 10s, and the growth rate of non-pattern even reached 300% (Fig. 8). And the results of circle-pattern showed that the ROS growth rate did not change significantly. Furthermore, analysis of the ROS growth rate showed a gradual decrease at 120s, confirming that CAP treatment induced an oxidative burst in more cells, leading to cancer cell death. 0s was used as a negative control for oxidative burst induction.

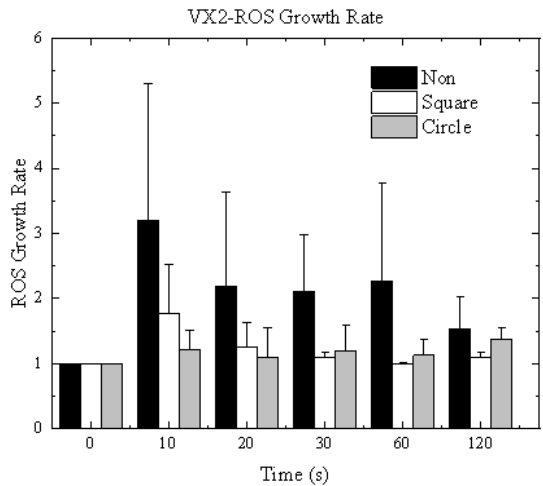

Fig. 8. ROS growth rate in air CAP with three patterns-treated VX2 cell lines.

As shown in Fig. 9a, in the case of air CAP with non-pattern, the ROS growth rate increased from control to 3min/30min incubation time and decreased later. In the case of square-pattern in Fig. 9b, the ROS growth rate slightly increased and then decreased for 10s and 20s treatment time, while there were no significant changes in ROS growth rate in other treatment times. For the circle-pattern in Fig. 9c, the content of ROS increased first and then decreased for 120s treatment time. The ROS growth rate for other treatment times did not show obvious changes. The figure showed that an increase in the conductive pattern increases superoxide anion production in cells and reaches the highest value in 30s.

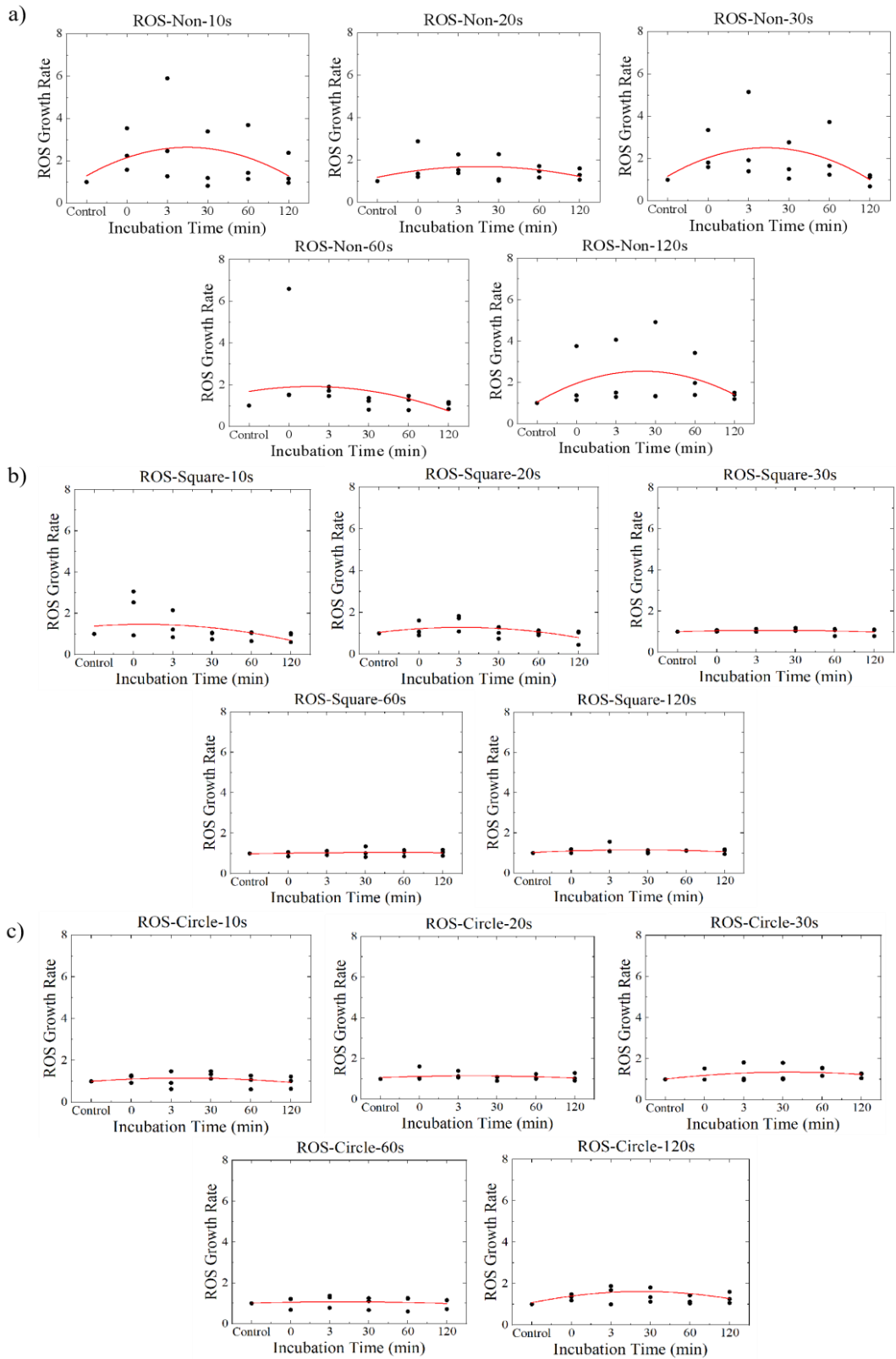

Fig. 9. ROS growth rate in air CAP with three patterns-treated VX2 cell lines with different incubation times (0min, 3min, 30min, 60min, and 120min) after different treatment times (10s, 20s, 30s, 60s, and 120s). a) ROS growth rate for air CAP with non-pattern. b) ROS growth rate for air CAP with square-pattern. c) ROS growth rate for air CAP with circle-pattern. Control represented VX2 cells without CAP treatment.

**3.3.3 Cell viability of VX2 cells after air CAP with patterns treatment**

The cell viability of CAP treatment time was investigated by evaluating CAP-induced cytotoxicity in VX2 cells. As shown in Fig. 10, VX2 cells were very sensitive to CAP exposure for more than 60s. With the increase in treatment time, the cell viability decreased. According to 10, 20, 30, 60, and 120s treatment time, the cell viability after 24h incubation decreased by about 29.5% (28.8%/1.2%), 65.2% (42.1%/20.7%), 65.4% (52.3%/27.7%), 37.0% (91.3%/75.8%), and 97.7% (97.1%/96.5%) for non-pattern (square-pattern/circle pattern), respectively. After 48h-incubation time, the cell viability decreased by about 12.6% (-14.0%/2.1%), 37.5% (57.1%/25.2%), 46.3% (64.5%/75.1%), 82.2% (89.8%/86.5%), 97.7% (97.4%/94.3%) for non-pattern (square-pattern/circle pattern), respectively.

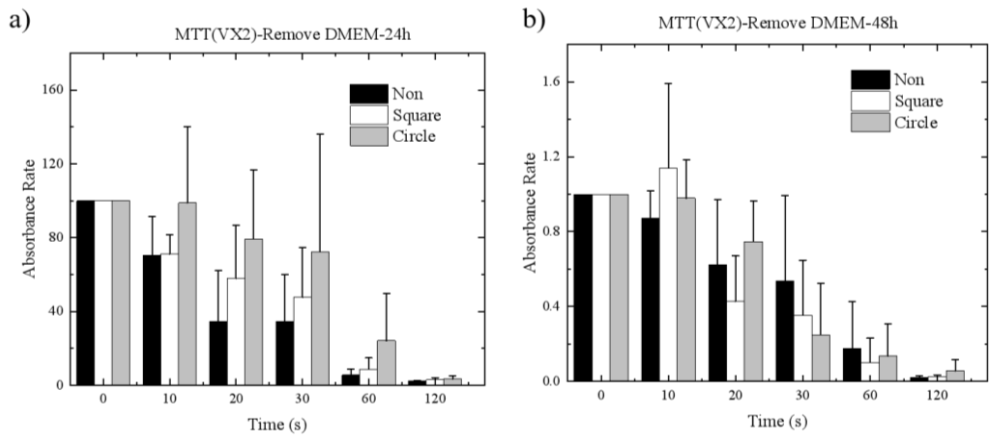

Fig. 10. Cell viability of VX2 cells after 24h (a) and 48h (b) of incubation with air CAP with three patterns of treatment (0s, 10s, 20s, 30s, 60s, and 120s treatment time).

## 4. Discussion

In recent years, scientists have tried to use CAP to replace or supplement drugs and traditional medical methods to achieve better treatment results[28-30]. Plasma medicine has made many research achievements in the application fields of wound healing, coagulation, tissue sterilization, root canal treatment, and cancer cell treatment[31-35]. However, CAP treatment of tumors has been hampered by the limitation of the gases necessary for CAP generation (mostly inert gases). Therefore, we aimed to develop and investigate the effects of air CAP with three different patterns of treatment on anaplastic squamous cell carcinoma (VX2 cell line) in vitro.

In this study, we designed and fabricated air CAP with three different patterns to investigate the plasma-induced changes in the concentration of active substances in VX2 cells and cultured medium under different treatment times, as well as their effects on cell viability. Circle-pattern had the shortest metal electrode circumference and square-pattern had the longest one. Circle-pattern had higher peak-peak discharge voltage and frequency than square-pattern and non-pattern. And the temperature variation of circle-pattern and square-pattern was lower than that of non-pattern. The OES measured by circle-pattern was lower than that of non-pattern and square-pattern, with the highest OES at the center of each air CAP with patterns. According to the electron energy distribution function (EEDF), the electron density of the circle-pattern was larger than that of the other two patterns. The experimental results show that DMEM can continuously store plasma-induced active substances, and CAP treated cells will release active substances into DMEM, which can be stored in DMEM within 24 hours. Cell viability decreased with the increased treatment time of CAP, and VX2 cells were inactivated in 120s. The cell inactivation ability of air CAP with non-pattern was the largest during the 24h-incubation time, while the cell extinguishing ability of circle-pattern was the largest for the incubation time of 48h.

Generally, plasma-generated ROS and RNS, including short- and long-life species resulted in cell death. We employed air CAP with different patterns to directly treat VX2 cells and exhibited a great effect in killing cancers cells (Fig. 10). Thus, we considered that short-life species played more important roles than long-life species. Short-lived radicals or species include hydroxyl radical (•OH), superoxide ($O_2^-$), nitrite (NO), peroxynitrite ($ONOO^-$), atomic oxygen (O), ozone ($O_3$), singlet delta oxygen (SOD, $O_2(^1\Delta g)$), and so on[36-40]. From Fig. 4 and Fig. 5, air CAP with different patterns generated lots of short-life species, which directly induced VX2 cell death. •OH-derived amino acid peroxides can contribute to cell injury because the •OH itself and protein peroxides can react with DNA, thereby inducing various forms of damage[41, 42]. $O_2^-$ is able to activate mitochondrial-mediated apoptosis via changing mitochondrial membrane potential resulting in cell death[43]. In addition, NO and $O_2^-$ are easy to form $ONOO^-$ once then collide or even locate within a few cell diameters, which is a powerful oxidant and nitrating agent being much more damage to tumor cells[44-46]. O and $O_3$ are known to have a strongly aggressive effect on cells[47, 48]. $O_2(^1\Delta g)$ produces oxidative damage and it selectively kills tumor cells in the emerging cancer therapy[49]. In addition, we utilized air CAP with patterns to activate cultured media, and our devices could generate much higher ROS and RNS concentrations (Fig. 6a) than other inert gas plasma devices. We also measured ROS and RNS concentration generated in VX2 cells after air CAP with patterns to show that plasma directly treated cells resulted in ROS/RNS concentration increasing in cells to induce death (Fig. 7, Fig.8, and Fig.9). Our results also indicated that VX2 cells after treatment of air CAP with different patterns released ROS and RNS to cultured media (Fig. 6b and 6c). Overall, the above results and discussion suggest that air CAP with patterns will be potential for cancer clinical applications. Further understanding of the exact underlying mechanisms will help determine the optimal combination as a therapeutic strategy.

## 5. Conclusions

We have developed air CAP with three patterns, which will generate uniform and stable plasma. Their active areas can be adjusted according to the size of the dielectric plate. The air CAP with three different patterns treated anaplastic squamous cell carcinoma (VX2 cell line) in vitro to show that they generated short-life species playing a more important role in inducing cell death. The air CAP with circle-pattern generated more active substances during at 60s treatment time, which resulted in a higher death rate of VX2 cells. Air CAP with three patterns was found to generate much higher concentrations of ROS and RNS than other inert gas plasma devices. When directly applied to cells, the air CAP with three patterns caused an increase in ROS/RNS concentration, leading to cell death. Our results also indicated that VX2 cells after treatment of air CAP with different patterns released ROS and RNS to cultured media. Overall, the air CAP with three patterns is a promising option for cancer clinical applications due to its convenience and potential effectiveness.

## Acknowledgments

This work was supported by the National Key R&D Program of China (2022YFE0126000, to Z.C.), the Guangdong Basic and Applied Basic Research Foundation (2022A1515011129, to Z.C.), and Ministry of Innovative Development of the Republic of Uzbekistan (grants number AL59-21122141, J.R).